\documentclass[12pt,preprint]{aastex}

\def\gsim { \lower .75ex \hbox{$\sim$} \llap{\raise .27ex \hbox{$>$}} }
\def\lsim { \lower .75ex \hbox{$\sim$} \llap{\raise .27ex \hbox{$<$}} }

\def\lesssim{\mathrel{\hbox{\rlap{\hbox{\lower4pt\hbox{$\sim$}}}\hbox{$<$}}}}
\def\gtrsim{\mathrel{\hbox{\rlap{\hbox{\lower4pt\hbox{$\sim$}}}\hbox{$>$}}}}
 
\shorttitle{M31 Minor Axis Profile}
\shortauthors{Irwin et al.}

\begin{document}
\slugcomment{ }
\title{A Minor Axis Surface Brightness Profile for M31\footnotemark[1]}
\footnotetext[1]{Based on observations made with the Isaac
  NewtonTelescope operated on the Island of La Palma by the Isaac
  Newton Group in the Spanish Observatorio del Roque de los Muchachos
  of the Instituto de Astrofisica de Canarias}
\author{Mike J. Irwin\altaffilmark{1}, Annette M. N. Ferguson\altaffilmark{2}, Rodrigo A. Ibata\altaffilmark{3}, Geraint F. Lewis\altaffilmark{4}, Nial R. Tanvir\altaffilmark{5}}
\altaffiltext{1}{Institute of Astronomy, Madingley Road, Cambridge UK CB3 0HA; mike@ast.cam.ac.uk}
\altaffiltext{2}{Institute for Astronomy, University of Edinburgh, Blackford Hill, Edinburgh UK EH9 3HJ; ferguson@roe.ac.uk}
\altaffiltext{3}{Observatoire de Strasbourg, 11, rue de l'Universit\'{e}, F-67000, Strasbourg, France; ibata@newb6.u-strasbg.fr}
\altaffiltext{4}{Institute of Astronomy, School of Physics, A29, University of Sydney, NSW 2006, Australia; gfl@physics.usyd.edu.au}
\altaffiltext{5}{Centre for Astrophysics Research, University of Hertfordshire, College Lane, Hatfield UK AL10 9AB; nrt@star.herts.ac.uk}

\begin{abstract}
  
  We use data from the Isaac Newton Telescope {\it Wide Field Camera}
  survey of M31 to determine the surface brightness profile of M31
  along the south-east minor axis.  We combine surface photometry and
  faint red giant branch star counts to trace the profile from the
  innermost regions out to a projected radius of 4\arcdeg\
  ($\approx55$~kpc) where $\mu_V\sim32$ mag~arcsec$^{-2}$; this is the
  first time the M31 minor axis profile has been mapped over such a
  large radial distance using a single dataset.  We confirm the
  finding by \cite{pritch94} that the minor axis profile can be
  described by a single de Vaucouleurs law out to a projected radius
  of 1.4\arcdeg\ or $\approx20$~kpc.  Beyond this, the surface
  brightness profile flattens considerably and is consistent with
  either a power-law of index $\sim-2.3$ or an exponential of
  scalelength 14~kpc. The fraction of the total M31 luminosity
  contained in this component is $\approx2.5\%$.  While it is tempting
  to associate this outer component with a true Population II halo in
  M31, we find that the mean colour of the stellar population remains
  roughly constant at V-{\it i} $ \approx1.6$ from
  $0.5-3.5$\arcdeg\ along the minor axis.  This result suggests that
  the same metal-rich stellar population dominates both structural
  components.  \end{abstract}


\section{Introduction}

The structure of galaxies at very large galactocentric radii (and
hence very faint light levels) is a subject of much current interest.
From a theoretical perspective, high resolution simulations of galaxy
formation indicate that many signatures of the galaxy assembly process
should lie buried in these parts (e.g. \cite{bull01,bull04}).  The discovery
of stellar substructure in galaxy halos (e.g.
\cite{maj03,ibata01,ferg02,shang98}) lends strong support to the idea
that galaxies assemble, at least in part, via the accretion of small
satellite systems.  However, equally important constraints on galaxy
formation come from studies of their smooth stellar components with
key questions concerning the profile, shape and composition of
stellar halos and outer disks.  Unfortunately, the low surface
brightness of these parts (typically $\mu \gtrsim
28-29$~mag~arcsec$^{-2}$) poses a significant challenge for studies of
diffuse unresolved light and few robust constraints have been derived
to date from this technique (e.g. \cite{dalc02,morr94}).

Several recent studies have mapped galaxy structure to well below the
usual depths of traditional surface photometry by taking advantage of
unique datasets and/or innovative analysis techniques.  \cite{zib04a}
carefully rescale and stack over 1000 edge-on galaxies selected from
the Sloan Digital Sky Survey (SDSS) in order to detect emission as faint as 
$\mu_r \sim 31$mag~arcsec$^{-2}$.  They find evidence for the
presence of flattened red power-law (c/a$\sim 0.6, \rho \propto
r^{-3}$) stellar halos around disk galaxies.  \cite{zib04b} take
advantage of the unprecedented depth of the Hubble Ultra Deep Field to
study faint extraplanar emission around an edge-on galaxy at $z=0.32$
and find a similar power-law halo component dominating at very faint 
surface brightness levels.  On the other hand, deep star count
analyses of the low luminosity systems NGC~300 and M33 indicate the
presence of only a disk component at very large radii 
(\cite{jbh05}, Ferguson et al. in prep).

In this Letter, we exploit the Issac Newton Telescope {\it Wide-Field
Camera} survey of M31 to determine the surface brightness profile of
the galaxy along the minor axis.  Using the combination of traditional
surface photometry in highly-crowded regions and faint red giant
branch (RGB) star counts in diffuse regions, we are able to trace the
profile from the innermost regions of M31 out to a projected radius of
4\arcdeg\ or $\approx55$kpc.  The contiguous nature of the survey
provides excellent sampling of the profile at all radii and the
ability to distinguish local density enhancements in M31 from
fluctuations in background galaxy counts and the foreground
distribution of Galactic stars.  Previous results from our photometric
survey have been reported in \cite{ibata01} and \cite{ferg02}.

\section{Observations}

The Isaac Newton Telescope {\it Wide Field Camera} is a 4-chip EEV
4k$\times$2k CCD mosaic camera which images $\approx$0.29 square
degrees.  During the period Sept 2000--Jan 2004, we used this camera
to image 163 contiguous fields (corresponding to $\approx 40$ square
degrees) in the disk and halo of M31.  Our map covers an
elliptical region of semi-major (minor) axis 4(2.5)\arcdeg\ or
$\approx$55(34)~kpc, with an additional $\sim 10$ square degree
extension towards the south.

Images were taken in the Johnson V and Gunn $i$ bands
under mainly good atmospheric conditions and typical seeing better
than 1.2\arcsec.  The exposure time of 800-1000s per passband per
field allowed us to reach $i=23.5$, V$=24.5$ (S/N$\approx$5) which is
sufficient to detect individual red giant branch (RGB) stars to M$_V
\approx 0$ and main sequence stars to M$_V \approx -1$ at the distance
of M31.  Details of the survey observing strategy and pipeline
reductions are discussed in \cite{ibata01} and \cite{ferg02}.

Objects are classified as noise artifacts, galaxies, or stars
according to their morphology.   In the
outer halo fields, we typically detect equal numbers of resolved and
unresolved sources within the magnitude and color ranges of interest.  Of the
extended sources, approximately 20\% are $`$compact' in the sense of
being within the 3-5 sigma range of the stellar boundary in the
classification statistic and having an ellipticity of $<0.4$.  Making
the plausible assumption that half of these are genuinely extended, we expect that the
contamination due to mis-classified barely resolved field galaxies is
small and, in general, considerably less than 10\% of the total number of
detected sources.  

Due to the presence  of the Galactic Plane, foreground
contamination increases smoothly from $\approx$13000 stars
per square degree at the southern extremity of the survey to
$\approx$20000 stars per square degree at the northern extremity
(integrated over all magnitudes).  This foreground variation, coupled
with our current lack of suitable comparison fields uncontaminated by
M31, is the primary uncertainty in our star counts at very large radii.

\section{Constructing the Minor Axis Profile}

Figure \ref{fig1} shows the surface density of stellar-like sources
with magnitudes and colours consistent with being RGB stars at the distance
of M31.  White patches in the map are either areas contaminated by
saturated stars (and the nuclear region of M31) or are due to small
gaps in the survey coverage.  As reported in earlier papers
\citep{ibata01,ferg02,lewis04}, the INT WFC survey has revealed a
wealth of stellar substructure in the outskirts of M31.  Our focus
here, however, is the smooth structure of the galaxy's outskirts.

The sources shown in Figure \ref{fig1} are selected to have $21 < i <
22$ and $i > 26.85-2.85({\rm V}-i)$ and we refer to these as
red RGB stars.  We select blue RGB stars as stellar sources with
$20.5 < i < 22.5$ and $ 24.85-2.85({\rm V}-i) < i <
26.85-2.85({\rm V}-i)$.  The lower magnitude limits for both red and
blue RGB stars are conservatively set at roughly one magnitude
brighter than the survey 5-$\sigma$ detection threshold to mitigate
against the effects of varying completeness. If age is constant across
the survey area, the red and blue selection criteria isolate
metal-rich and metal-poor giants respectively.

Our large-area contiguous survey of M31 enables us to trace the
surface brightness profile along the southern minor axis from the
inner regions out to well beyond previously-published measurements
\citep{pritch94,durr04}.  The presence of significant substructure in
the outskirts of M31, especially in the southern half of the galaxy,
means the region over which to compute a representative profile must
be chosen carefully.  Inspection of the distribution of RGB stars
orthogonal to the minor axis indicates that a region of $\pm
0.5$\arcdeg\ on either side of the minor axis (defined as position
angle of 141.9\arcdeg) is free of obvious contaminating debris (see
Figure \ref{fig1}).  Debris from the giant stellar stream is a serious
problem beyond $\approx$0.5\arcdeg\ to the west of the minor axis,
whereas the north-east shelf overdensity is a problem further
eastward.

In order to study the light profile in the inner regions of M31, we
create a large mosaiced image from individual V- and $i$-band
pointings resampled at 1\arcsec\ resolution.  Overlaps between
pointings are used to adjust for sky brightness variations during the
observations.  We then carry out surface photometry on this mosaic.
The V- and {\it i}-band surface brightness is computed as
the median value in a series of equally-spaced (10\arcsec)
wedge-shaped bins with width increasing linearly from
100\arcsec\ in the innermost regions to 820\arcsec\ at a radius of
1\arcdeg, to ensure overlap with the star counts.  This sampling was
chosen as a compromise between maximising resolution in the inner
parts and signal-to-noise considerations in the outer parts, and
minimising the number of bright star halos affecting the surface
brightness photometry. This technique works well out to a radius of
$\approx 0.5\arcdeg$ beyond which uncertainties in the overall sky
level correction start to become significant.  Beyond 1.0\arcdeg, the sky
background dominates the signal and these regions are used to
determine the appropriate background value to subtract from the measured
surface brightness profile.

To explore the profile further out, we employ RGB star counts computed
in rectangular bins of size 3\arcmin $\times$ 1\arcdeg\ (see the
dashed lines in Figure \ref{fig1}).  The validity of using this
technique in conjunction with the direct surface brightness measures
rests on the fact that a significant fraction of the light of an old
stellar population originates in luminous RGB stars.  The RGB star
counts are split into blue and red RGB components and are used to
extend the V- and {\it i-}band profiles respectively.  We correct the star
counts for incompleteness due to crowding using the prescription of
Irwin \& Trimble (1984), although by restricting the use of star
counts to the outer regions of M31 the correction is never larger than
a factor of two.  An additional additive correction to the star counts
is made for ``background'' contamination arising from foreground Milky
Way stars and unresolved background galaxies.  This level is set by
computing the surface density of unresolved sources with magnitudes
and colours typical of M31 RGB stars in several of the outermost
pointings in our INT WFC survey, lying at radii 4--5\arcdeg.  We
assume that these sources are entirely contaminants and while this may
not be strictly correct due to a possibly very extended M31 halo
\citep{raja05}, it does not have a major impact on the profile over
the surface brightness range that we are considering here.

Finally, the star count profiles are shifted vertically\footnote{The exact shifts
applied are $39.0-2.5{\rm log}_{10}$(counts) and 
$37.4-2.5{\rm log}_{10}$(counts) to the blue and red star counts respectively.
In each case, $``$counts'' denotes the background-corrected RGB star counts
per square degree.} to overlap with the photometrically-calibrated inner surface brightness profiles
in the radial range $0.4-0.7$\arcdeg.  In joining the profiles in this
way, we implicitly assume that the slope of the luminosity function is
constant with radius; that is, that the RGB star counts represent a
constant fraction of the total flux at each position.  This assumption
seems valid given the lack of an obvious radial metallicity gradient
within the M31 halo (\cite{ferg02,durr04}, and see Section 4).  Figure
\ref{fig2} shows the combined V- and {\it i-}band minor axis profiles.
That the profiles are smooth across the overlap region serves as
reassurance that no features or slope changes are introduced as a
result of combining profiles constructed from the different
techniques.

\section{Results and Discussion}

Overlaid on the surface brightness profile in Figure \ref{fig2}
is the de Vaucouleurs R$^{1/4}$ law that \cite{pritch94} found
best-characterised their minor axis profile over the
radial range 1\arcsec\ to 1.5\arcdeg.   This model has an 
effective radius of r$_e=0.1\arcdeg$ or 1.4~kpc, derived from scaling
\cite{pritch94}'s original value to the currently-accepted distance of 785~kpc
\citep{mccon05,durr01}.   As can be seen, this model also provides
an excellent description of the profile derived here.  The slight
mismatch between the data and the model at radii of 0.1--0.4\arcdeg\
occurs where the contribution of the disk component -- which we have
neglected here -- to the minor axis profile is a maximum
\citep{wk88,pritch94}.

This result goes beyond merely confirming the findings of previous
studies.  Earlier  work has relied on measurements from various sources
in order to cover a sufficient radial range along the minor axis
\citep{pritch94}. These measurements have not only been made by
different authors with different telescopes/instruments, but often in
different filters, and several assumptions have been required to place
them on a common photometric system.  The uncertainty this has
introduced into the minor axis profiles published to date has been
unclear.  Our ability to determine the profile from 0.02 to 4\arcdeg\
using a single dataset thus represents a significant advance.

Beyond a radius of 1.4\arcdeg, the
minor axis profile levels off relative to the extrapolation of the inner
R$^{1/4}$ law.  Such a flattening has also recently been reported by
\cite{raja05} but at a slightly larger radius of 
of 2.2\arcdeg.  These authors refer to this outer component as the
$`$halo'.  Overlaid on the left-hand panel of Figure \ref{fig2} is a
projected NFW profile \citep{nfw97,yang03} of scale-radius
0.25\arcdeg\ or 3.4~kpc.  NFW profiles provide excellent descriptions
of the density profiles of dark matter halos and asymptote towards
$\rho \propto r^{-3}$ at large radii 
(though we note that an NFW dark matter halo would have a much larger
scale length than the above).  
Figure \ref{fig2} shows that while an
NFW profile with this scale radius provides a reasonably good
match to the outer parts of the minor axis profile, the radial
density fall-off is actually somewhat steeper than r$^{-3}$.  Indeed, a
power-law fit to the minor axis surface brightness profile beyond
1.5\arcdeg\ yields an index of $\approx -2.3$.  Alternatively, the
outer profile can also be adequately fit with an
exponential of scalelength 1\arcdeg\ or $\sim14$~kpc.

An advantage of NFW and exponential profiles is that they remain
finite when integrated from the centre (although in the case of an NFW
profile an outer radius limit is also required, which we adopt here as
100~kpc).  Using the parameters noted previously, we find that the
absolute V-band magnitude of the halo is -17.1 and -17.0 for an NFW
and exponential profile respectively. Allowing for 0.2 mag of
extinction and a total M31 V-band magnitude of M$_{V_0}=-21.1$
\citep{hodge94}, this implies the fraction of the total M31 luminosity
contained in the halo component is only $\approx 2.5\%$.

Figure \ref{fig3} shows a V-{\it i} minor axis colour profile
constructed from the individual V- and {\it i-}band surface brightness
profiles.  The colour reddens over the inner 0.5\arcdeg\ but then
levels off to a constant value of $\approx 1.6$, within the extent of
the uncertainties.  Rather remarkably, the break in the minor axis
profile at 1.4\arcdeg\ is unaccompanied by any obvious change in
colour, implying that the same metal-rich stellar population dominates
at all radii.  This is further supported by \cite{durr04}'s finding of
the same metallicity distribution function at 20 and 30~kpc, radii
which bridge the region where the surface brightness profile changes
slope.  Taken together, these results make it difficult to associate
the flat outer component seen along the minor axis of M31 with a
classical $`$Population II' halo.

\begin{center}
\begin{figure}
\includegraphics*[angle=0,scale=0.8]{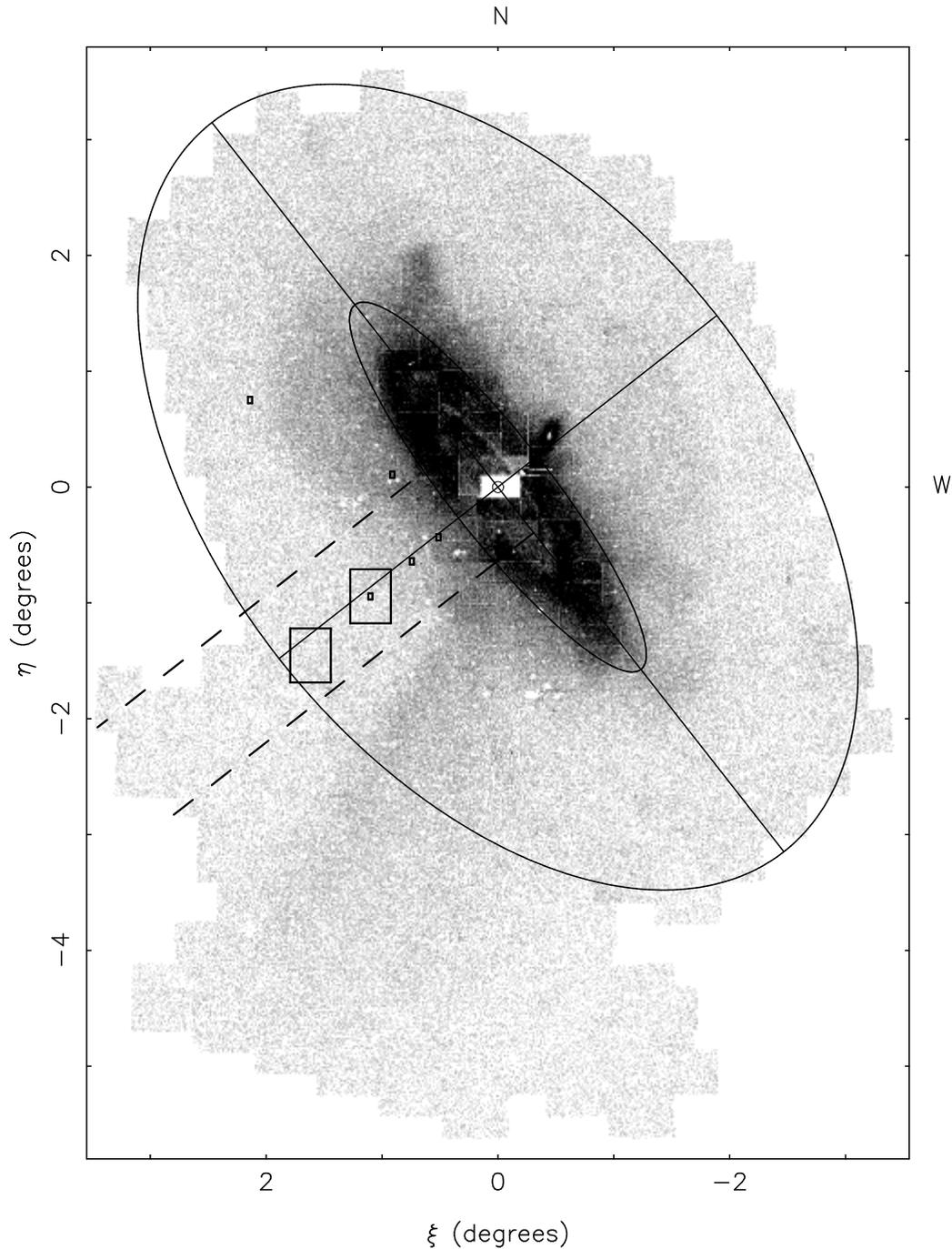}
\caption{A standard coordinate projection of the surface density of  
red RGB stars in a
$40$ square degree area around M31.  The outer
ellipse denotes a flattened ellipsoid (aspect ratio 3:5) of semi-major
axis 55~kpc, while the inner ellipse has a semi-major axis
27~kpc and represents the approximate extent of the bright disk.  The
area used to compute the minor axis surface brightness profile
($\pm0.5\arcdeg$ of the minor axis) is delineated by the dashed
lines. Overlaid are the locations of the fields observed by
\cite{durr01} and \cite{durr04} (large rectangles) and \cite{pritch94} (small 
rectangles).\label{fig1}}
\end{figure}
\end{center}

\begin{figure}[h]
\includegraphics*[angle=-90,scale=0.75]{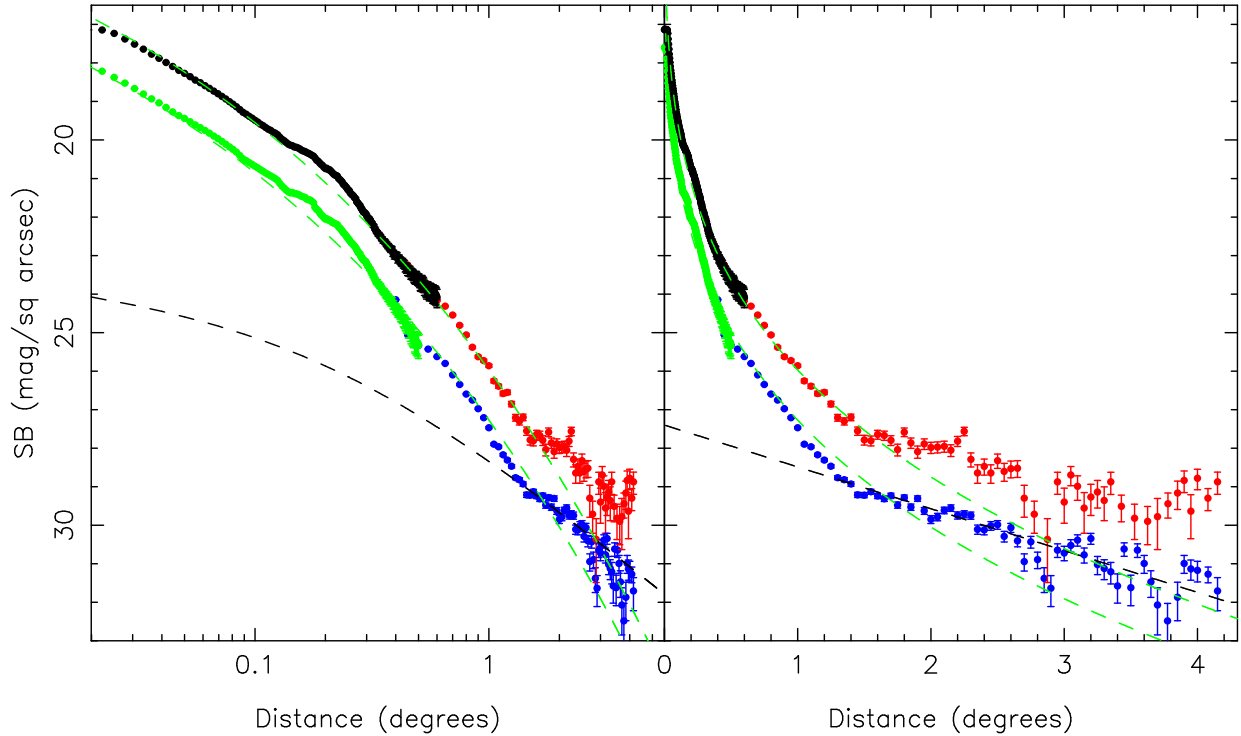} 
\caption{The effective V and {\it i}-band minor axis profiles shown on a
log-log (left) and log-linear (right) scale.  The V-band profile is
illustrated in green and blue and the {\it i}-band profile in black
and red. The green and black circles are derived from surface
photometry whereas the blue and red points are derived from star
counts in the magnitude and colour selection boxes described in the
text.  The error bars reflect a combination of poissonian and
background uncertainties. The green dashed lines show a de Vaucouleurs
R$^{1/4}$ law with $b_{eff}=1.4$~kpc. The dashed black line in the
left-hand panel shows an NFW profile computed with
scale-radius=3.4~kpc and, in the right-hand panel, an exponential
profile computed with scalelength=13.7~kpc. \label{fig2}}
\end{figure}

\begin{center}
\begin{figure}
\includegraphics*[angle=-90,scale=0.8]{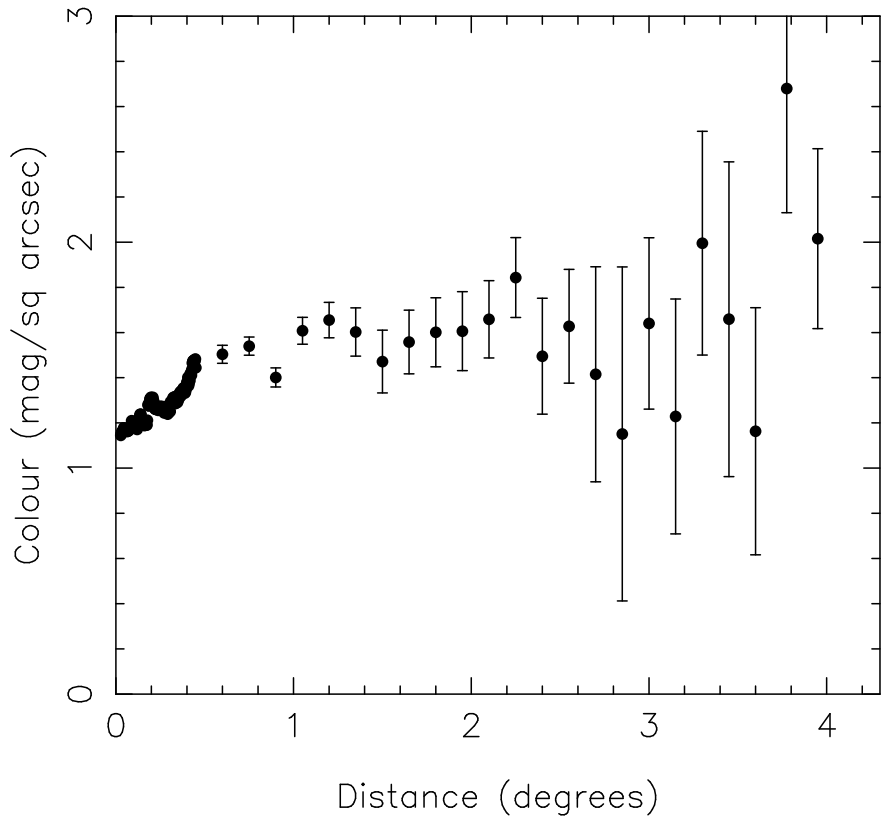} 
\caption{A minor axis colour profile constructed from the effective
 V and {\it i}-band profiles in Figure \ref{fig2}.  The star count
data are binned up into 0.2\arcdeg\ bins; error bars reflect uncertainties in
the individual profiles as well as the dispersion within bins. \label{fig3}}
\end{figure}
\end{center}

Understanding the nature of this new structural component in M31 will
require further photometric and spectroscopic studies of its
constituent populations. For example, do these stars exhibit a
distinct kinematical signature from the rotating inner components
\citep{ibata05}? Does the halo shape remain constant with increasing
radius? \cite{pritch94} find no evidence for the halo becoming
significantly rounder out to $\sim 2\arcdeg$ however inspection of
Figure \ref{fig1} indicates that the off-axis fields they used to
measure the flattening would have been contaminated by stellar debris.

Our finding of a red power-law halo component at large radius in
M31 is especially interesting in light of the \cite{zib04a} and
\cite{zib04b} analyses of faint extraplanar emission around distant
galaxies. The stellar halos observed in those studies fall off as $\rho
\propto r^{-3}$ to $r^{-4}$, similar to the $\rho \propto r^{-3.3}$
found here. These indices also compare favourably to that of the Milky
Way halo
\citep{prest91,kinman94,ivezic00}.  Additional studies of the faint 
outskirts of  galaxies
will be required to understand whether these power-law
components are indeed ubiquitous and to derive constraints on their
origin.

\end{document}